# Disentangling the Effects of Simultaneous Environmental Variables on Perovskite Synthesis and Device Performance *via* Interpretable Machine Learning

Tianran Liu[1,2]*, Nicky Evans[1,2,3]*, Kangyu Ji[1,2]*, Ronaldo Lee[3], Aaron Zhu[3], Vinn Nguyen[1], James Serdy[1], Elizabeth Wall[4], Yongli Lu[5], Florian A. Formica[6], Moungi G. Bawendi[5], Quinn C. Burlingame[4], Yueh-Lin Loo[4], Vladimir Bulovic[3], Tonio Buonassisi[1]

[1]Department of Mechanical Engineering, Massachusetts Institute of Technology, Cambridge, Massachusetts 02139, USA

[2]Research Laboratory for Electronics, Massachusetts Institute of Technology, Cambridge, Massachusetts 02139, USA

[3]Department of Electrical Engineering and Computer Science, Massachusetts Institute of Technology, Cambridge, Massachusetts 02139, USA

[4]Department of Chemical and Biological Engineering, Princeton University, Princeton, New Jersey 08544, USA

[5]Department of Chemistry, Massachusetts Institute of Technology, Cambridge, Massachusetts 02139, USA

[6]Atinary Technologies Inc., Route de la Corniche 4, 1066 Epalinges, Switzerland

*These authors contributed equally

Email: liutianran1121@gmail.com; nickye17@mit.edu; axvcb1597382@gmail.com; buonassi@mit.edu

## Abstract

Despite the rapid rise in perovskite solar cell efficiency, poor reproducibility remains a major barrier to commercialization. Film crystallization and device performance are highly sensitive to environmental factors during fabrication, yet their complex interactions are not well understood. In this work, we present a systematic framework to investigate the influence of both individual and coupled environmental variables on device efficiency and crystallization kinetics. We developed an integrated fabrication platform with precise, independent control over ambient solvent partial pressure, absolute humidity, and temperature during spin-coating and thermal-annealing processes, respectively. Using the platform, we implemented a closed-loop Bayesian optimization framework to efficiently explore the multi-dimensional processing space. We mapped the impact of these environmental variables on device performance and identified coupled effects among them. *In-situ* grazing-incidence wide-angle X-ray scattering measurements further validated these findings by revealing a nonlinear interaction between absolute humidity and solvent partial pressure during spin-coating, which affects crystallization dynamics. To isolate and quantify these interactions, we developed an interpretable machine learning approach that combines knowledge distillation with Shapley interaction analysis. The model revealed the contribution of each interaction varies across different processing conditions. Our study highlights the importance of integrated ambient sensing and control to achieve repeatable perovskite solar cells, and demonstrates the utility of combining active learning with interpretable machine learning to navigate complex, high-dimensional processing landscapes.

**Introduction**

Metal halide perovskite solar cells (PSCs) have attracted significant attention due to their high power conversion efficiencies (*PCE*s), potential for low-cost manufacturing, and compatibility for integration into tandem photovoltaic architectures.[1–7] Despite these advantages, the industrial-scale deployment of PSCs is impeded by pronounced inter- and intra-sample variability, which stems in part from an incomplete understanding of the underlying physicochemical mechanisms.[8–10] Variability in perovskite crystallization kinetics and resulting device performance arises from both known and uncontrolled processing factors.[11–13] These introduce nuisance variables that undermine sample-to-sample consistency, limit recipe reproducibility across laboratories, and hinder the transfer of protocols from research settings to scalable manufacturing platforms.

Previous studies have identified a broad range of deposition variable classes that influence perovskite crystallization kinetics, and consequently, device performance.[14–16] These include precursor-related parameters (*e.g.*, chemical composition,[17] solvent system,[18] molarity,[19] precursor age,[20] and purity[21,22]), process-related factors (*e.g.*, deposition rate,[23,24] film-formation dynamics,[25] and annealing conditions[26]), and environmental conditions (*e.g.*, ambient temperature during deposition,[27] water vapor,[28–30] and potentially partial pressures of solvents[31]). Among these, environmental variables are often the most difficult and expensive to control, and thus despite their central importance, they are frequently under-characterized, under-reported, or entirely overlooked in both academic research and industrial practice. This deficit is widely regarded as a key contributor to the irreproducibility of perovskite device fabrication.[8,32–34]

Existing investigations into environmental effects typically rely on one-variable-at-a-time experimental frameworks. However, given the potential for coupled (nonlinear) interactions among environmental factors, there is a need to elucidate the governing process-structure-property relationships within a multivariate framework. In the absence of such understanding (and corresponding capabilities for real-time monitoring and control), even modest fluctuations in ambient conditions may result in irreproducible perovskite film properties and erratic device performance across even small-area solar cells, effects that are further amplified during scale-up, leading to non-uniformities across large-area coatings.

In this study, we developed and utilized a fully integrated fabrication platform comprising two custom-built environmental chambers—one for spin coating and one for thermal annealing. This system enables independent and precise control over key environmental parameters, including solvent partial pressure (SPP), absolute humidity (AH), and temperature, during the deposition and annealing of $FAPbI_3$ perovskite films. Using this platform, we implemented a closed-loop Bayesian optimization (BO) framework to efficiently explore the multi-dimensional processing space and identify optimal environmental conditions. By reviewing the multi-dimensional search space of our machine learning model, we uncovered nonlinear interactions among environmental variables that significantly influence device efficiency. These findings were supported by *in-situ* grazing-incidence X-ray diffraction (GIWAXS), which captured distinct crystallization kinetics that were nonlinearly affected by AH and SPP during spin coating. To quantify how environmental variables interact, we developed an interpretable machine learning approach that integrates knowledge distillation and Shapley interaction analysis. The results show how changes in one processing condition can compensate for or counteract the effects of another, revealing the complex contribution of each interaction across different conditions. Our study highlights the importance of precise environmental control in achieving reproducible perovskite devices and demonstrates how the integration of active learning and interpretable machine learning facilitates efficient exploration and deep understanding of complex, high-dimensional processing spaces.

**Results**

*a. Framework for multivariate environmental analysis*

We first constructed our environmental fabrication platform comprising two isolated environmental enclosures—the spin-coating (SC) chamber and the thermal-annealing (TA) chamber, **Figure 1a** and **Supplementary Figure 1**) connected via a transport channel. This setup enables precise, independent measurement, and control of key environmental variables during each stage of $FAPbI_3$ perovskite film fabrication. Each enclosure features an independently controllable environment, equipped with a set of sensors and control systems that regulate environmental parameters according to user-defined setpoints. In the SC chamber, sensors measure AH and chamber temperature, while the partial pressure of dimethylformamide (DMF) vapor is measured using a dedicated air pump that draws air from the vicinity of the substrate stage and delivers it to a solvent vapor sensor. The AH and DMF vapor partial pressure can be actively controlled by conditioning the incoming air through inputs of compressed dry air (CDA), a DMF bubbler, and a humidifier. The AH can be further increased by using a water boiler to supply water vapor. A resistive heater is used to raise the chamber temperature, while a fan circulates the atmosphere to ensure uniform environmental conditions throughout the chamber. In the TA enclosure, sensors monitor AH and temperature in a similar fashion. Only AH is actively regulated by adjusting the flows of CDA and humidified air as well as the water boiler. Temperature control is not implemented, as the high temperatures generated by the hotplate dominate the thermal environment within the enclosure. All sensors are positioned near the substrate to ensure that the recorded conditions closely reflect those experienced by the perovskite films during processing. We note that AH is used as the environmental parameter instead of relative humidity (RH), as RH depends on temperature. To isolate and study independent environmental effects, AH offers a decoupled representation of water vapor content.

As illustrated in **Figure 1b**, our workflow begins with perovskite film fabrication inside the two custom-built isolated environmental chambers. The perovskite films were fabricated using a one-step antisolvent method, followed by an annealing process. We defined a multi-dimensional search space (Step 1) comprising key environmental parameters including AH, temperature, and SPP. Five environmental parameters were actively monitored in real time: AH, temperature, and SPP of DMF vapor in the SC box, as well as AH and temperature in the TA box. Of these, four parameters (all except the TA box temperature) were actively controlled during fabrication. Devices were fabricated based on an *n-i-p* perovskite architecture: $FTO/TiO_2$/perovskite/Spiro-OMeTAD/Au. $TiO_2$, pre-deposited by the supplier *via* chemical bath deposition (CBD), was used as the electron transport layer (ETL) across a large batch. The high reproducibility of these $FTO/TiO_2$ substrates ensures higher signal-to-noise ratio when investigating environmental effects during perovskite film formation.[35] The devices were subsequently tested for photovoltaic (PV) performance, and the resulting data (including their environmental conditions and device metrics) were compiled into a centralized database. In Step 2, a Gaussian process-based surrogate model was iteratively updated using experimental results, allowing the system to learn the performance landscape. Based on the surrogate model and an acquisition function, the algorithm selects the next experimental condition (Step 3) to explore the search space, forming a closed-loop learning cycle that accelerates optimization. After multiple iterations, we performed knowledge distillation on the BO model (Step 4), in which a simplified student model was trained to approximate the behavior of the surrogate model. In Step 5, a Shapley interaction analysis was performed on the student model to quantify the nonlinear and synergistic effects of the environmental variables on device performance, enabling post hoc root-cause analysis.

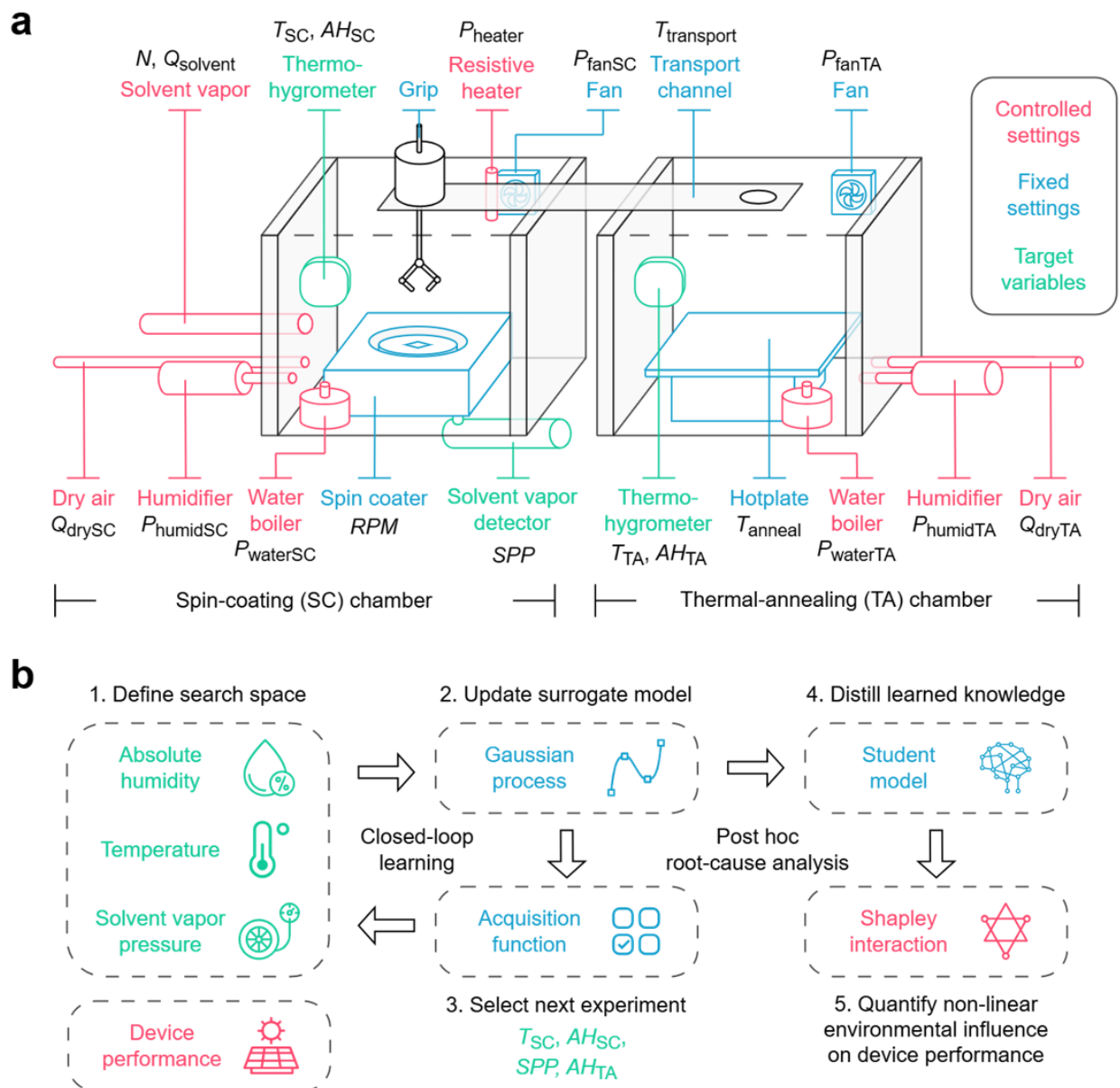

**Figure 1. Environmental fabrication platform and active learning workflow. (a)** Environmental fabrication platform comprising two isolated environmental enclosures: the spin-coating chamber and the thermal-annealing chamber. **(b)** Closed-loop active learning framework for optimizing environmental conditions and interpreting their impact on device performance.

We validated the effectiveness of our environmental enclosures by fabricating a batch of eight devices (Sample IDs 1–8 in **Supplementary Table 1**) under baseline conditions: both chambers were continuously purged with CDA to maintain minimal AH, and the SC box was kept at ambient temperature. After each sample, the chamber was purged until the SPP returned to zero. All devices were fabricated using the same nominal input conditions. The results (**Supplementary Figure 2**) show the absolute values for max-min variation (and as a relative percentage of the mean value of the dataset) for each device parameter fabricated with environmental control as 0.02 V (1.9 % rel.) in open-circuit voltage ($V_{OC}$), 0.90 mA/cm$^2$ (3.6 % rel.) in short-circuit current density ($J_{SC}$), 3.89 % (5.2 % rel.) in fill factor (*FF*), and 1.04 % (5.2 % rel.) in *PCE*. These correspond to standard deviations of 0.009 V, 0.26 mA/cm$^2$, 1.40 % and 0.35 %, respectively, quantifying the reproducibility under controlled environmental conditions in our enclosures.

To minimize the risk of other variables affecting repeatability, our device-fabrication protocol was modified as follows: (1) all films were fabricated by one single experienced researcher, (2) an observer verified consistency in processing steps, (3) the same precursor chemistry was used throughout, and (4) all powders were sourced from the same batch.

To quantify how simultaneous environmental variables affect device performance, we adopted a BO framework. The input variables included DMF vapor partial pressure, AH, and temperature in the SC enclosure, as well as AH in the TA chamber. The sole output variable was the *PCE* of the resulting devices, measured under standard 1-sun illumination. The experimental campaign consisted of five rounds. **Figure 2a** and **Supplementary Table 2** summarize the parameter space explored across all five rounds, underscoring the multivariate nature of the study.

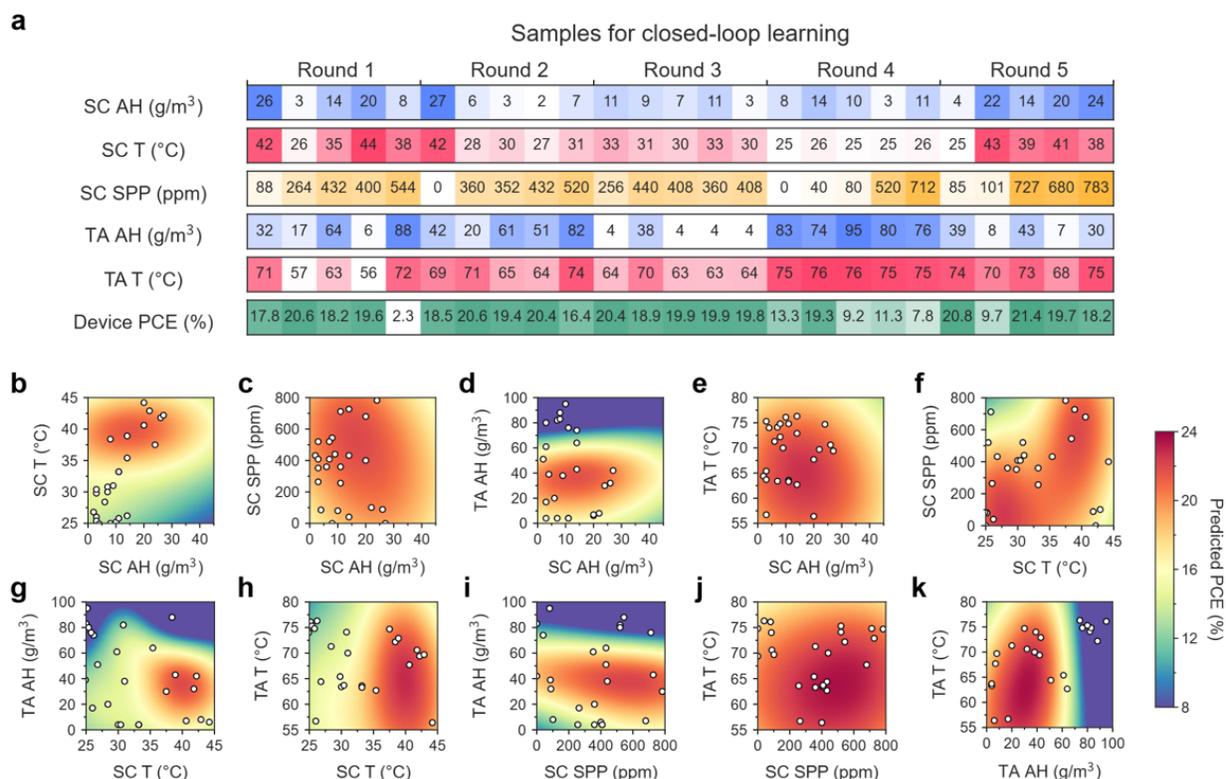

**Figure 2**. **Environmental variables in perovskite film fabrication and their impact on device efficiency.** (**a**) Summary of experimental input variables (rows 1–5) and an output variable (row 6), across 5 cycles of learning (one initial sampling round, and four active-learning rounds). SC AH: spin-coating absolute humidity; SC T: spin-coating chamber temperature; SC SPP: spin-coating chamber solvent partial pressure; TA AH: thermal-annealing chamber absolute humidity;

TA T: thermal-annealing chamber temperature. (**b-k**) Plots showing joint influences of input parameters on efficiency. The color map indicates the average value predicted by the fitted Gaussian process regressor. Prediction uncertainty is shown in **Supplementary Figure 3**.

The campaign began with an initial sampling round ("Round 1") comprising five experiments, generated using a custom Latin hypercube sampling (LHS) strategy that integrated Clausius-Clapeyron constraints and a diversity-enforcement scheme appropriate for low-data regimes (see **Methods**), to ensure uniform coverage of the search space. Initial sampling and data visualization scripts were developed in-house and are available in the **Data availability** section. This was followed by four active learning cycles ("Rounds 2–5"), each with a target batch size of five experiments, executed using Atinary's no-code Bayesian optimization platform. Their proprietary optimizer "Falcon GPBO" leverages a Gaussian process as surrogate model, well-suited to systems exhibiting continuous input–output behavior, and an acquisition policy to explore the combinatorial space defined by all possible experimental designs. At each cycle, the surrogate model was re-fitted on dataset augmented with the acquired data from the previous round, and the acquisition policy proposed experimental conditions to be tested, thus completing this iteration of the BO loop.

To visualize the optimization landscape and enable human supervision, we generated plots showing the expected mean predictions of the Gaussian process across pairwise parameter combinations, with overlays indicating sampled and suggested points. **Figures 2b–k** show the pairwise plots at the end of the experiment (after Round 5). Notice the Clausius-Clapeyron relation is respected in **Figures 2b** and **2k**. The contours (mean device *PCE*) represent predicted values from our custom-coded Gaussian process regressor used for data visualization, and do not represent actual efficiencies. The parity plot showing the correlation between predicted and actual efficiencies, indicating the quality of the model used for data visualization, is shown in **Supplementary Figure 4**.

*b. Limits of marginal analysis and evidence of interactions*

In systems with multiple input variables, single-variable effects are often evaluated using Shapley feature importance ranking. One advantage of this game-theoretic approach is that it fairly distributes interaction effects among features, avoiding the attribution bias seen in sequential methods like waterfall charts, where interactions may be disproportionately assigned to the first variables considered. SHapley Additive exPlanations (SHAP) is a specific implementation to explain the outputs of a trained regressor.

As shown in **Figure 3**, we applied SHAP to examine how individual environmental input variables affect predicted *PCE* in the trained Gaussian process regressor. In order of decreasing priority, the top three environmental variables affecting efficiency globally (across the entire dataset) are: AH of the thermal annealing step, temperature of the spin coating step, and solvent vapor partial pressure of the spin coating step.

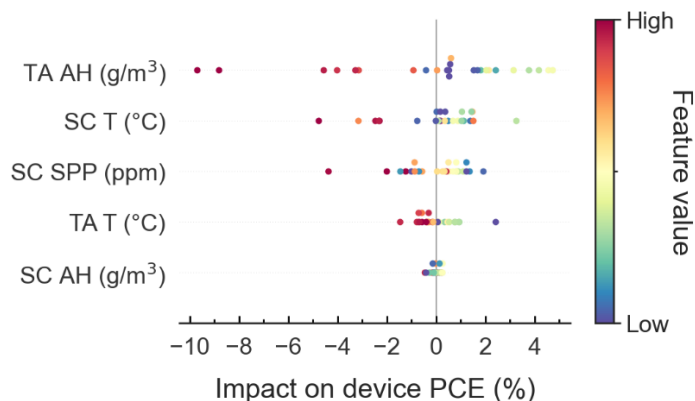

**Figure 3. SHAP summary plot ranking the relative influence of environmental variables on perovskite *PCE*s.** The SHAP values are derived from a Gaussian process regressor trained on experimental data from this study (**Supplementary Table 2**).

We note that while SHAP provides valuable insights into feature importance, it is not sufficient on its own to fully capture the effects of environmental variables on *PCE*. One limitation is that feature importance, as quantified by SHAP, can vary significantly across different regions of the input parameter space. This variability can stem from feature interactions (coupling between variables), which SHAP does not explicitly quantify. Indeed, we find that the assumption of linear superposition of environmental variables does not always hold in our study. This is reflected in the concave or convex shapes of the plots in **Figure 2**, which result from nonlinear joint contributions of two environmental variables, whereas a purely additive contribution is expected to produce a flat or rectilinear contour plane.

We observed that SHAP analysis identifies the variable "Spin coater solvent partial pressure," as one of the most influential individual features affecting *PCE* (**Figure 3**). However, at high solvent vapor partial pressures, both high and low efficiencies are observed (**Supplementary Table 2**), suggesting the presence of more complex interactions that cannot be captured through marginal importance alone. When we applied a tree-based regression model (**Supplementary Figure 5**), we observed that DMF vapor partial pressure during spin-coating only negatively affects efficiency when absolute humidity is high. This suggests that non-linear interactions are present between water, solvent molecules, and the perovskite precursor film as it dries and crystallizes. This observation is supported by selected literature reports.[36]

To further investigate these cross-variable interactions, we utilized an environmental enclosure at beamline 11-BM of the National Synchrotron Light Source II (NSLS-II) to collect *in-situ* GIWAXS while spin-coating and dripping anti-solvent onto FAPbI$_3$ perovskite precursor films in humidity- and DMF-vapor-controlled atmospheres. The enclosure was maintained at ambient temperature (~22 °C) and continuously flushed with N$_2$ streams passing through independently-controlled bubblers containing water and DMF to control the humidity and DMF vapor partial pressure. Measurements were collected over a 90-second window following the onset of spin coating, with an anti-solvent dispensed after approximately 23 s. Four conditions were examined, corresponding to combinations of 0 g/m³ and 3.9 g/m³ AH (0% and 20% RH at 22 °C) with 0 ppm and 800 ppm DMF, enabling a direct comparison of individual and coupled environmental effects on the structural evolution of the film. (Please note that these values indicate the amount of added water or solvent vapor to the ambient gas.)

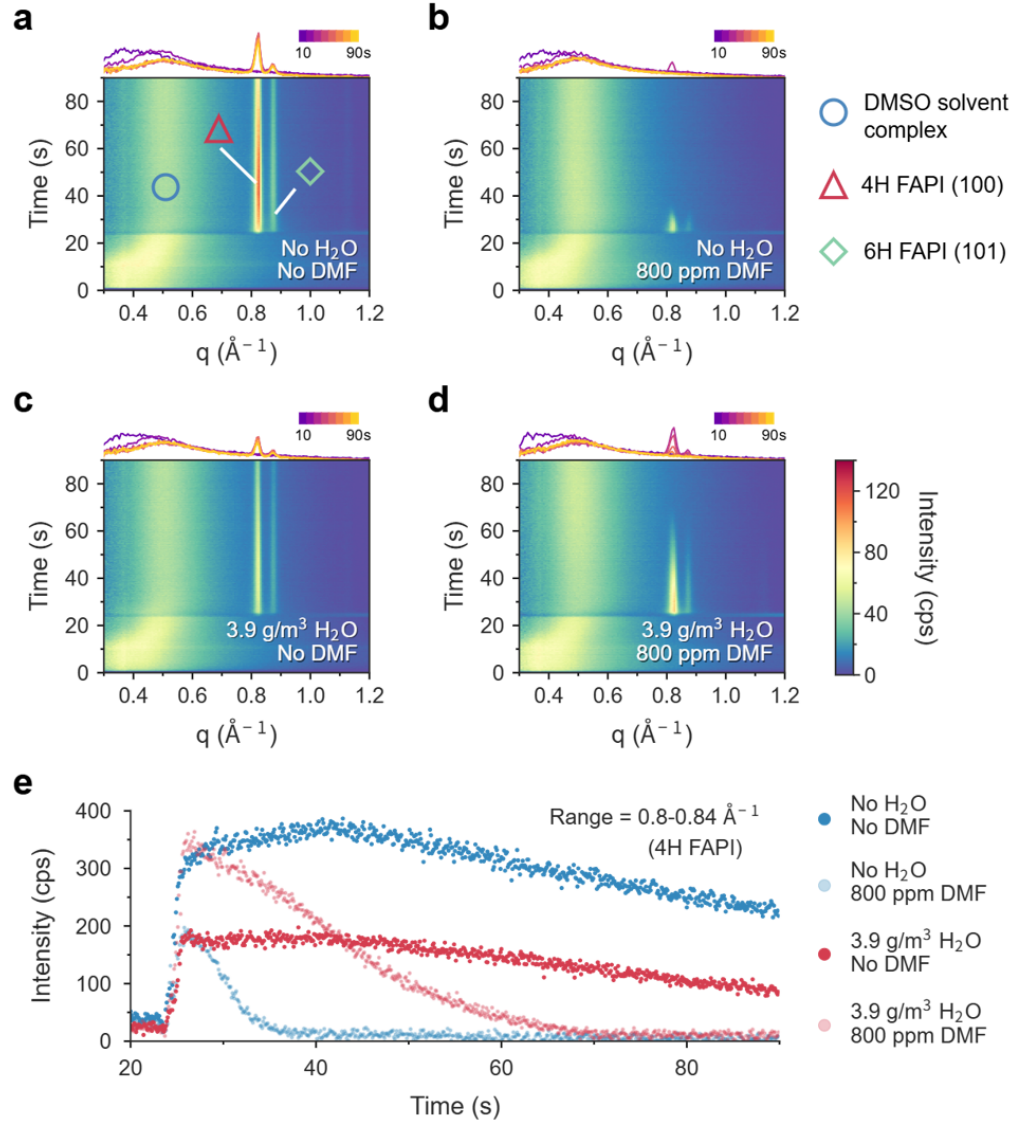

**Figure 4. GIWAXS analysis reveals non-additive effects of DMF vapor and humidity on perovskite crystallization dynamics.** (a–d) Time-resolved GIWAXS maps showing the evolution of perovskite crystallization under different environmental conditions, with added water or DMF vapor as indicated. Characteristic scattering features correspond to the DMSO–PbI$_2$ complex (○), 4H FAPI (100) (△), and 6H FAPI (101) (◇). (e) Integrated intensity of the 4H FAPI (100) peak over time across the four conditions in (a-d).

The GIWAXS diffractograms are shown in **Figure 4**. The broad scattering feature observed prior to dispensing the anti-solvent arises from the disordered precursor solution. As the solvent-saturated film thins and DMF begins to evaporate, this broad feature gradually shifts to higher $q$-values, where it stabilizes into a solvent-complex (FA$_x$MA$_{2-x}$Pb$_3$I$_8$·2DMSO), which appears consistently across all conditions studied. After the antisolvent is dispensed (causing the horizontal discontinuity in all spectra at ~23 s), the intensity of the FA$_x$MA$_{2-x}$Pb$_3$I$_8$·2DMSO feature is reduced and peaks corresponding to the hexagonal 4H and 6H phases of FAPbI$_3$ emerge. These phases persist throughout the measurement under solvent-free environments, though their intensity is notably reduced under the elevated 3.9 g/m$^3$ humidity condition.

To better visualize phase evolution, absolute intensities of the 4H-FAPbI$_3$ phase across the four conditions are plotted in **Figure 4e**. In a DMF-rich atmosphere with no added humidity, the 4H and 6H peaks rapidly decrease and disappear within 10 s. As humidity increases to 3.9 g/m$^3$ under the same DMF-rich atmosphere, the peaks fade, but persist for more than 30 s. Although the precise mechanism requires further study, it is possible that DMF vapor may slow the evaporation of solvent from the film, *e.g.*, *via* Le Chatalier's principle, causing these phases to redissolve into solvent complexes. Interestingly, we do not observe evidence of α-FAPbI$_3$ or the δ-FAPbI$_3$ (2H) phase during spin-coating in any of the atmospheres tested here. Thus we infer that the nucleation and growth of α-FAPbI$_3$ occurs entirely during annealing.

Overall, these results highlight that both DMF vapor partial pressure and humidity significantly influence film formation. However, their effects are not simply additive. When altered simultaneously, they result in qualitatively different outcomes compared to changes made independently. This exemplifies the complex, non-linear nature of environmental interactions during spin coating. A more targeted study will be needed to elucidate the mechanisms of combined effects on perovskite crystallization pathways.

*c. Framework to analyze coupled environmental effects on efficiency*

To address the limitations of marginal feature analysis, we decoupled the nonlinear feature interactions between environmental variables by quantifying Shapley interactions through the calculation of both joint and individual marginal contributions of variable pairs.[37] We note that the Gaussian process regressor from BO is governed by kernel function and thus not explicitly modeled, leading to difficulty in disentangling feature pair effects. To address this, we employed multiple student models to distill the knowledge obtained from the Gaussian regressor teacher model. These student models include Support Vector Regression (SVR), Decision Tree Regression (DTR), Random Forest (RF), and Multi-Layer Perceptron (MLP) (see 5-fold cross validation in **Supplementary Figure 6**). We visualize typical feature interactions, for example recipes using network plots (**Figure 5** and **Supplementary Figure 7**), where each node represents an environmental variable and each edge corresponds to the interaction between the connected nodes. We find that the contribution of each edge varies across recipes. For example, the interaction between DMF vapor partial pressure and temperature during spin coating can have either a negative, negligible, or positive contribution on the final *PCE* (**Figure 5**, **b-d**). This means that a standard trial-and-error approach, where a single parameter is optimized while keeping others fixed, may not yield the best halide perovskite film for devices, as the effects of optimization cannot simply be summed up, and nonlinear effects exist in such complex scenarios (see dependence plots of predicted *PCE* as a function of perovskite fabrication parameters in **Supplementary Figure 8**). Moreover, our results indicate that multiple sub-optimal processing conditions can lead to similar device performance, highlighting the need for multivariable optimization strategies.

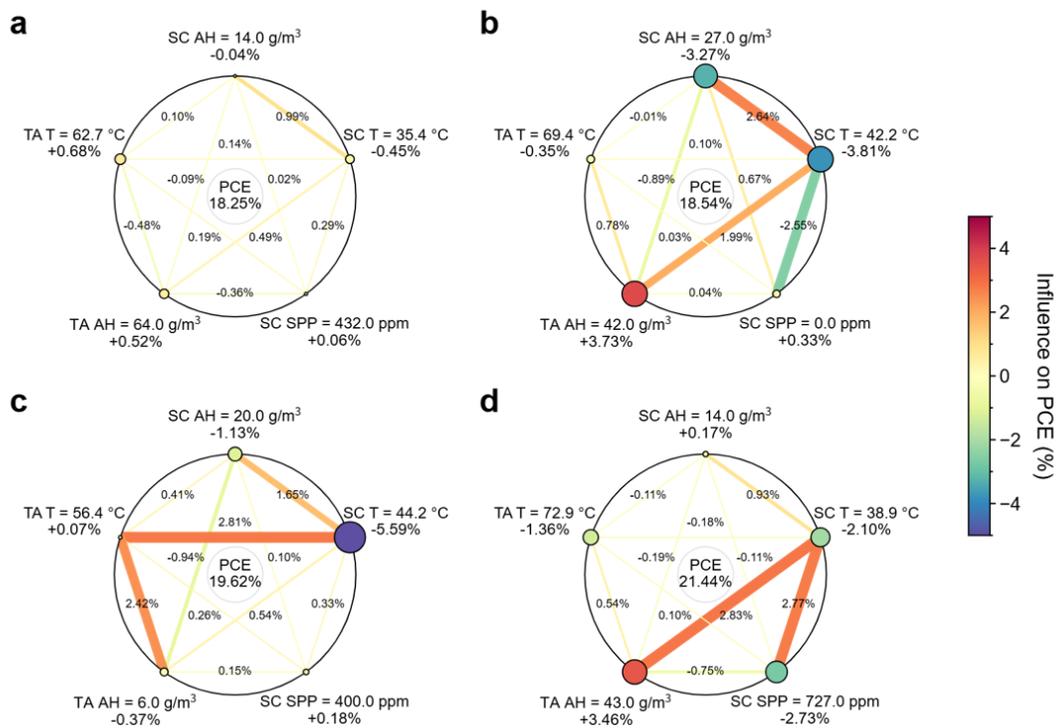

**Figure 5**. **Nonlinear feature interactions in model predictions of *PCE*s for four selected devices fabricated in different conditions.** Each feature (environmental variable) is represented as a node. The size of each node reflects the strength (Shapley value) of its individual contribution to the device *PCE*, with larger nodes indicating stronger main effects. The strength and direction of interactions between features are depicted by the color and thickness of the edges, where thicker edges represent stronger nonlinear interactions. The final device *PCE* is calculated by summing the contributions from all nodes and edges, starting from a baseline *PCE* of 15.46%.

We note that calculating and interpreting higher-order interactions among multiple environmental variables is challenging. However, based on our findings, we hypothesize that such higher-order interactions exist among different environmental parameters during perovskite film fabrication. We note that such higher-order interactions can be captured by a well-trained regressor, highlighting the value of machine-learning optimization strategies such as BO to expediently navigate complex environmental spaces and identify the global optimum. This also highlights the need for precise sensing and control of environmental variables, to ensure consistent and optimized halide perovskite device fabrication.

**Discussion**

We demonstrate that environmental variables (specifically: AH, temperature, and SPP) can exert significant and sometimes coupled effects on the device efficiency, crystallization kinetics, and repeatability of $FAPbI_3$ PSCs. By actively controlling these variables during spin coating and annealing, we isolate their individual and joint effects on device performance. For the specific one-step $FAPbI_3$ deposition process used here, higher humidity levels and the presence of non-zero DMF vapor partial pressure notably reduces performance, revealing a nonlinear interaction between solvent environment and moisture content. While we expect that specific feature importances and processing windows will vary across compositions and fabrication protocols, the methodology presented here provides a generalizable framework for fingerprinting a given recipe's environmental sensitivities. Such insights may ultimately aid in evaluating recipe robustness and manufacturability, particularly for scale-up to large-area, high-throughput production. More broadly, our results highlight the necessity of precise measurement and control of environmental factors, not only to optimize small-area device performance, but also to enable reproducible and uniform manufacturing of perovskite modules at scale.

**Author contributions:** TL, JS, and TB conceived and designed the study. JS, TL, VN, AZ, NE, RL developed the enclosure. TL, NE, RL, YL fabricated and characterized devices. TL, NE, RL, AZ, KJ, FF, and TB used or wrote the code. EW performed the synchrotron-based GIWAXS measurements and analysis. KJ, TL, NE, EW, and TB produced the figures, and TL, KJ, NE, and TB wrote the manuscript. MB, QB, YLL, VB, and TB supervised the research. All authors contributed to reviewing, editing, and approving the manuscript.

**Data availability:** Raw data and code is available on: https://github.com/PV-Lab/perovskite_environmental_control

**Acknowledgements:** We thank Armi Tiihonen for discussion about constrained optimization, Lucien Brey and Riccardo Barbano for their assistance with BO, and the entire ADDEPT team for feedback throughout the project, especially the team of David P. Fenning including Maimur Hossain and Daniel Abdoue. This material is based upon work supported by the U.S. Department of Energy's Office of Energy Efficiency and Renewable Energy (EERE) under the Solar Energy Technologies Office Award Number DE-EE0010503. EW, QB, and YLL thank Dr. Ruipeng Li and Alan Kaplan for their help with X-ray scattering measurements, which were conducted at the Center for Functional Nanomaterials (CFN) and the Complex Materials Scattering (CMS) Beamline 11-BM of the National Synchrotron Light Source II (NSLS-II), which are both US DOE Office of Science Facilities, at Brookhaven National Laboratory under Contract no. DE-SC0012704.

**Conflicts of interest:** Some of the authors (TL, NE, TB, YLL, VB, QB, EW) are co-authors on patents and patent applications concerning perovskite photovoltaics. QB has a financial stake in Rayleigh Solar Tech, Inc. Atinary Technologies Inc. is a for-profit company. No conflicts are deemed to exist with this study, and all code and data is publicly available.